\documentclass[12pt,preprint]{aastex} 
\usepackage{bm}
\usepackage{amsmath}
\usepackage{float}
 
\shorttitle{} 
\shortauthors{} 
 
\begin{document} 
 
\received{} 
\accepted{} 
 
\title{Gravity Modes Reveal the Internal Rotation of a Post-mass Transfer Gamma Doradus/Delta Scuti Hybrid Pulsator in Kepler Eclipsing Binary KIC~9592855}  
 
\author{Z. Guo$^{1,2}$, D. R. Gies$^{2}$, R. A. Matson$^{3,2}$} 

\affil{$1.$ Copernicus Astronomical Center, Polish Academy of Sciences,  Bartycka 18, 00-716 Warsaw, Poland;
guo@camk.edu.pl }

\affil{$2.$ Center for High Angular Resolution Astronomy and  
 Department of Physics and Astronomy, 
 Georgia State University, P. O. Box 5060, Atlanta, GA 30302-5060, USA;
gies@chara.gsu.edu, rmatson@chara.gsu.edu} 

\affil{$3.$ NASA Ames Research Center, Moffett Field, CA 94035, USA}

\slugcomment{10/11/2017} 

 
\begin{abstract} 

We report the discovery of a post-mass transfer Gamma Doradus/Delta Scuti hybrid pulsator in the eclipsing binary KIC~9592855. This binary has a circular orbit, an orbital period of 1.2 days, and contains two stars of almost identical masses ($M_1=1.72M_{\odot}, M_2=1.71M_{\odot}$).
However, the cooler secondary star is more evolved ($R_2=1.96R_{\odot}$) while the hotter primary is still on the zero-age-main-sequence ($R_1=1.53R_{\odot}$). Coeval models from single star evolution cannot explain the observed masses and radii, and binary evolution with mass-transfer needs to be invoked. After subtracting the binary light curve, the Fourier spectrum shows low-order pressure-mode pulsations, and more dominantly, a cluster of low-frequency gravity modes at about $2$ day$^{-1}$. These g-modes are nearly equally-spaced in period, and the period spacing pattern has a negative slope. We identify these g-modes as prograde dipole modes and find that they stem from the secondary star.  The frequency range of unstable p-modes also agrees with that of the secondary. We derive the internal rotation rate of the convective core and the asymptotic period spacing from the observed g-modes. The resulting values suggest that the core and envelope rotate nearly uniformly, i.e., their rotation rates are both similar to the orbital frequency of this synchronized binary.

\end{abstract} 
 
\floatplacement{figure}{H}
 
 

\section{Introduction}

Pulsating stars that show gravity modes are great assets to stellar astrophysics, because we can sound the deep interior through asteroseismology. Detailed analysis of their Fourier spectrum, especially for those on the main-sequence (Slowly Pulsating B stars and $\gamma$ Doradus stars), only became available quite recently (Zwintz et al.\ 2017; P{\'a}pics et al.\ 2015, 2017; Van Reeth et al.\ 2015a), thanks to the continuous observations from space missions (e.g., {\it Kepler} : Borucki et al.\ 2010; {\it MOST} : Walker et al.\ 2003). Among these stars, the $\gamma$ Dor pulsators, which are early-F to late-A type stars with $M \in [1.4M_{\odot}, 2.0M_{\odot}]$, are very common and have gained importance in the recent literature (Bedding et al.\ 2015; Van Reeth et al.\ 2015b, 2016; Schmid et al.\ 2016). Most of the well-studied $\gamma$ Dor stars are single stars or they reside in very detached binaries, which can be treated as the result of single star evolution. 
Binary star evolution with mass transfer can also form $\delta$ Scuti and $\gamma$ Dor variables (Chen et al. 2017), as well as other pulsating stars (subdwarf B stars: Vos et al.\ 2015; RR Lyraes and Cepheids: Karczmarek et al.\ 2017 and Gautschy \& Saio 2017). Asteroseismology of these abnormal $\delta$ Scuti and $\gamma$ Dor pulsators has not been carried out, and most of the past studies focused on the frequency analysis, e.g., oscillating Algol (oEA) systems (Mkrtichian 2002, 2003). In a previous paper (Guo et al.\ 2017), we have shown that the high-frequency p-modes observed in a post-mass transfer $\delta$ Scuti star in the eclipsing binary KIC~8262223 agree with the excited range from the non-adiabatic calculations, albeit with some minor differences.
In this work, we derive the accurate fundamental parameters of a post-mass transfer $\gamma$ Dor/$\delta$ Scuti hybrid in the eclipsing binary (EB) KIC~9592855 and decipher its g-mode pulsations.
Those double-lined eclipsing binaries containing g-mode pulsators are ideal targets  to refine our theory of stellar structure and evolution. For a list of 16 $\gamma$ Dor pulsating EBs, please refer to \c{C}ak{\i}rl{\i} et al.\ (2017).

KIC~9592855\footnote{2MASS J19350483+4614117, ASAS J193505+4614.2}  ($\alpha_{2000}$=$19$:$35$:$04.833$, $\delta_{2000}$=$+46$:$14$:$11.70$,  $K_p=12.216$, $V=12.255$) was first classified as a detached Algol-type eclipsing binary (Al*) in the All Sky Automated Survey (Pigulski et al.\ 2009). It was included in the {\it Kepler} Eclipsing Binary Catalog (Pr{\v s}a et al.\ 2011; Slawson et al.\ 2011). A set of light curve parameters is estimated from neural networks, including temperature ratio ($T_2/T_1=0.97$), sum of relative radius ($R_1/a+R_2/a=0.489$), eccentricity and argument of periastron ($e \sin \omega=0.0011$, $e \cos \omega=0.0089$), and orbital inclination ($\sin i=0.961$). The latest updates of the catalog are described in Kirk et al.\ (2016).

Debosscher (2011) did an automatic search for variabilities in the first Quarter light curves of {\it Kepler}, including KIC~9592855. The main frequencies they found are all orbital harmonics: $1.6402$ day$^{-1}$, $8.2010$ day$^{-1}$, and $9.8412$ day$^{-1}$, which are 2, 10, and 12 times of orbital frequency, respectively. Thus, these frequencies only indicate the binary nature, and the real pulsation frequencies are buried in the binary signal. 
Gies et al.\ (2012) measured the eclipse times  of this binary, yielding an orbital period of $P=1.21932475(2)$ days and the time of primary minimum of $T_0=2455656.3029(1)$ in Barycentric Julian Date (BJD), where $1\sigma$ uncertainties are given in parentheses in units of the last digits. Gies et al.\ updated their measurements using all 17 Quarters of {\it Kepler} data in Gies et al.\ (2015), and the updated ephemeris is  $P=1.21932487(2)$ and $T_0=2455691.664073(7)$. 
Conroy et al.\ (2014) also measured the eclipse timing variations of this system, together with 1278 short-period {\it Kepler} eclipsing binaries. The overall consensus is that there is no signature of a third companion to this binary, and the $O-C$ residuals scatter around zero with an amplitude of about $50$ seconds. Throughout this work, we use the term `primary' for the star that is in eclipse at the time of the deeper minimum (primary minimum) $T_0$.
    
The latest photometry-derived atmospheric parameters reported by the {\it Kepler} team (Huber et al.\ 2014) are those based on Brown et al.\ (2011): $T_{\rm eff}=7513\pm 262$K,  $\log g=3.946\pm 0.400$, ${\rm [Fe/H]}=-0.015\pm 0.300$ and those updated in Pinsonneault et al.\ (2012): $(T_{\rm eff}, \log g, {\rm [Fe/H]} )=(7498, 3.964, -0.060)$. A mass of $1.73M_{\odot}$ is derived from fitting the Dartmouth isochrones (Dotter et al.\ 2008) in the $\log g- T_{\rm eff}$ plane. The binary spectral energy distribution (SED) fitting of Armstrong et al.\ (2014) yields effective temperatures of the primary and secondary star as $7934\pm 381$K and $7805\pm 570$K, respectively. 

This article is organized as follows. In Section 2, we describe how the photometric and spectroscopic data are analyzed. It includes characterizing the atmospheric properties of individual components and detailed binary modeling. Section 3 details the interpretation of the pulsation spectrum, a general comparison between observations and the theoretically unstable modes, the identification of a series of prograde dipole g-modes and how the internal rotation is derived, and in the end the evolutionary history of this binary. In the final section (Sec. 4), we summarize our results and discuss prospects for future work.
\section{Observations and Binary Modeling}

\subsection{Spectral Characteristics of Individual Components}
As one of the systems from our sample of 41 {\it Kepler} eclipsing binaries in a {\it Kepler} GO program (Gies et al.\ 2012), KIC~9592855 was observed with the R–C Spectrograph mounted on the 4 meter Mayall telescope at the Kitt Peak National Observatory (KPNO). We obtained seven spectra with moderate resolving power ($R=\lambda/\delta \lambda \approx 6000$) in the wavelength range of $3930-4610$\AA.  The signal-to-noise ratio (S/N) of the spectra ranges from $70$ to $120$.

We show one of the observed double-lined spectra in Figure 1 (green line). At the orbital phase of 0.18, the two components are clearly separated in the cores of the Balmer lines (H$\delta\lambda4102$ and H$\gamma\lambda4341$). We measured the Doppler shifts of the spectral lines by cross-correlating the observed spectra with templates from the UVBLUE library (Rodr{\'{\i}}guez-Merino et al.\ 2005). The detailed data reduction and cross-correlation procedures can be found in Matson et al.\ (2016). The derived radial velocities for this target together with the other 40 {\it Kepler} eclipsing binaries listed in Gies et al.\ (2012, 2015) will be presented in a separate paper (Matson et al.\ 2017).

We implemented the Doppler tomography algorithm (Bagnuolo \& Gies 1991) to get the individual spectrum of each component.  In this algorithm, the linear inverse problem of spectral separation was solved with an iterative method in the wavelength domain. Other methods include those based on Singular Value Decomposition (Hadrava 1995; Ilijic 2004) and  MCMC (Czekala et al.\ 2017).
We determined the mean flux ratio of the two stars in the observed spectral range ($\approx 4225$ \AA) as $F_2/F_1=1.00\pm 0.04$.
The separated individual spectrum was compared with a grid of synthetic spectra from the UVBLUE library to determine the atmospheric parameters: $T_{\rm eff}, \log g, v \sin i,$ and [Fe/H] following the method detailed in Guo et al.\ (2016). In brief, the method includes fixing the value of surface gravity $\log g$ to those from the ELC binary modeling and optimizing the remaining parameters in a chi-square-minimization sense with both the genetic algorithm {\it PIKAIA} (Charbonneau 1995) and the gradient-based Levenberg-Marquardt algorithm\footnote{The MPFIT package by Craig B. Markwardt}. The $1\sigma$ uncertainties were determined by varying the parameters so that $\chi^2$ increased by 1.0 from the minimum value in the former algorithm, and from the local covariance matrix in the latter algorithm. The results of the two methods agree very well and are summarized in Table 1. Note that the reported errors are only formal and likely underestimated. The factors that could affect the errors include the systematic errors in the normalization of the spectra and the grid interpolation errors when using the synthetic templates.

In Figure 1, we display the reconstructed individual spectrum of each star (black) and their best-fitting models (red). The spectra of the two components have similar overall appearances, although the derived effective temperature of the primary star is higher ($T_{\rm eff1}=7567\pm 65$K, $T_{\rm eff2}=7037\pm 65$K). The ratio of the eclipse depth in the {\it Kepler} light curve also indicates that the secondary star is slightly cooler than the primary. The two stars have very similar projected rotational velocities ($(v \sin i)_1=61\pm 10$ km s$^{-1}$, ($v \sin i)_2=62\pm 10$ km s$^{-1}$).  Both components have essentially a solar metallicity within $\sim 1 \sigma$ ([Fe/H]$_1=-0.05\pm 0.04$ and [Fe/H]$_2=-0.05\pm 0.04$). Note that resolution of our spectra in velocity space is 26.25 km s$^{-1}$ per pixel, and
we are unable to reliably measure small rotational velocities
($v \sin i < 30$ km s$^{−1}$).

Given the relatively low rotational velocities compared with single $\delta$ Scuti stars and the temperatures we have derived, these two stars may be metallic line or Am stars.  These often show stronger \ion{Sr}{2} $\lambda 4077$ and weaker
  \ion{Ca}{1} $\lambda 4226$ lines compared to normal dwarfs (Gray \& Corbally
  2009), and both differences appear to be present in a comparison of the
  reconstructed and model spectra of the primary and secondary (Fig.~1).
  However, the Am classification depends critically on the relative strengths of the \ion{Ca}{2} K-line and metallic lines, and unfortunately,
  the \ion{Ca}{2} $\lambda\lambda 3933,3968$ lines are too blue to be recorded on our spectra.  Spectra with greater blue coverage are needed to investigate this issue further. A detailed analysis of element abundances by using high-resolution spectra is also desirable.

\subsection{{\it Kepler} Photometry and Binary Modeling}

Broadband photometric data of KIC~9592855 with micro-magnitude precision were collected by the {\it Kepler} satellite from Q1 to Q17. We use the available long cadence Simple
Aperture Photometry (SAP) light curves with a sampling rate of 29.44 minutes. The light curves were obtained from the Mikulski Archive for Space Telescopes (MAST) and prepared following several procedures ($5\sigma$ clipping of outliers, removing of low-frequency trend with splines, and median value corrections, etc.) detailed in Guo et al.\ (2016). We adopted a mean contamination factor\footnote{ ELC corrects the effect of aperture contamination of flux from nearby stars by adding to the median value of the model light curve $y_{med}$ an offset $ky_{med}/(1- k)$.} of $k=0.004$ over all quarters.

The light curve synthesis code ELC (Orosz \& Hauschildt 2000), originally designed to model X-ray binaries, can also be used to model detached/semi-detached eclipsing binaries, non-eclipsing binaries, and transit light curves of exoplanets. It is used to model the binary light curves (LCs) and radial velocities (RVs) of KIC~9592855 simultaneously. The synthetic LCs and RVs are calculated from integrating the specific intensities from the NextGen model atmosphere (Hauschildt et al.\ 1999) over the stellar surface described by the Roche model. We assume a circular orbit with synchronized rotations as suggested by spectroscopy. Due to the sharp differences between the quality and quantity of LCs and RVs (micro-magnitude vs. km s$^{-1}$ precision, $\sim 50000$ LC data points vs. 7 RV measurements), we have to scale the errors of RVs so that RVs and LCs have comparable contributions to the total chi-squares. The genetic algorithm PIKAIA is implemented to find the best combination of parameters. The fitting parameters include those that impact the light curves ($i, f_1,f_2,T_{\rm eff2},l_1,l_2$) and those that mainly affect the radial velocities ($K_1,q,\gamma$). The orbital eccentricity ($e$) is fixed to zero, and the gravity darkening coefficients ($\beta$) are fixed to the canonical value $0.08$ associated with the stellar envelope properties (convective). The detailed definitions of all the parameters can be found in Table 2. The orbital period ($P$) and the time of primary minimum ($T_0$) are fixed to values derived by Gies et al.\ (2015). These ephemeris parameters are of high precision, and experiments show that letting them vary does not improve the light curve fit. 

For eclipsing binaries, the temperature ratio can be derived from the relative depths of the eclipses. In the modeling, we usually fix one temperature to break the strong degeneracy between the two temperautres. We fixed the effective temperature of the primary star to a grid of values ($T_{\rm eff1}\equiv 7100$ $,7200,7300,7400,7500,7600$ K) and found that the derived temperature ratio changes only slightly ($T_{\rm eff2}/T_{\rm eff1}=0.987\pm 0.010$). This suggests the two stars have very similar temperatures and the difference inferred from the light curve is about $100$ K.
However, the values of spectroscopic temperatures of the two stars are different by $\sim 500 $ K as listed in Table 1. This discrepancy ($\sim 400$ K ) between the light curve and spectroscopic values of effective temperatures also exists in other studies, e.g., the secondary star in KIC~3858884 has $T_{\rm eff2}=6890$ K from the spectroscopic analysis which is $\sim 300$ K higher than that from the light curve (Maceroni et al.\ 2014). To reconcile the above discrepancy, we adopted the temperature solutions that are close to the mean value of both spectroscopic temperatures $T_{\rm eff} ({\rm mean})=7302$ K since the temperature ratio is very close to one. The final adopted effective temperatures are $T_{\rm eff1}\equiv 7300$ K and $T_{\rm eff2}=7202$ K\footnote{Note our derived effective temperatures are lower than the literature values which are estimated from the photometry (see Section 1). Those literature values do not take into account the binarity. We also checked the  Ca I 4226 and Fe I 4271 lines which are good temperature diagnostics for such stars (Gray \& Corbally 2009). The comparison with UVBLUE models indeed confirms that the literature values are overestimated.}. Note that different fixed values of $T_{\rm eff1}$ do not have a significant effect on other binary parameters.
Our final binary model is shown in Figure 2, and the associated ELC parameters are listed in Table 2. Our binary model can fit the LCs and RVs very well. The model suggests the primary and secondary star have almost identical masses ($M_1=1.72M_{\odot}, M_2=1.71M_{\odot}$) but very different radii ($R_1=1.53R_{\odot}, R_2=1.96R_{\odot}$). The implications of these measurements on the evolutionary history are presented in Section 3.4. The $v \sin i$ value of the primary star from our synchronized binary model agrees with the spectroscopic one within $1 \sigma$, but the secondary $v \sin i$ only agrees within $\sim 1.5\sigma$. 

The light curve residuals still show sinusoidal variations with an amplitude of $\sim 0.002$ mag, and they arise from an imperfect match of ellipsoidal variations. The sinusoidal shape over the whole four-year dataset implies that they are not due to spots or stellar activities.
We tried to change the reflection parameter, gravitational darkening coefficient, and even the rotation period, but
this mismatch persists. We conclude that it is possibly due to the limitations of the ELC model or the optimization algorithm. Indeed, the high-precision of {\it Kepler} light curves have called for the need for a refinement of our binary modeling tools (e.g., Pr{\v s}a et al.\ 2013).

\section{Asteroseismology and Binary Evolution}

\subsection{Pulsational Properties}

We subtracted the best-fitting binary light curve from the original data and did a Fourier analysis of the pulsation residuals. There are still some systematic residuals in the eclipses (Figure 2). These residuals appear sinusoidal and they generate some aliases of orbital harmonics of low amplitudes in the Fourier spectrum. To remedy this, we generate a simple binary light curve model by binning the phase-folded light curve. This model can fit the data equally well, but the pulsation residuals have much better quality in the eclipses. We thus proceed in our analysis using these residuals. 

The Fourier spectrum was calculated to the Nyquist frequency of {\it Kepler} long cadence data ($f_{\rm Nyquist}$=24.46 d$^{-1}$). The significant frequencies (using the classical criterion of the signal to noise ratio $S/N > 4$) were extracted with the pre-whitening procedures implemented with the {\it Period 04} package (Lenz \& Breger 2005). These extracted frequencies are enumerated in the order of decreasing $S/N$, and they are listed in Table 3 and 4.
The uncertainties are calculated following the treatment in Kallinger et al.\ (2008). In Figure 3, we show a section of the pulsation residuals and the corresponding Fourier amplitude spectrum.

In the pulsation spectrum, the dominant pulsational frequencies cluster around $\sim 2$ day$^{-1}$, and the strongest pulsation mode is at $2.2326$ day$^{-1}$. Close inspection reveals that these low-frequency pulsations belong to a series of prograde dipole g-modes. The detailed analysis will be presented in Section 3.3, and the parameters of these pulsation modes are listed in Table 3.

There are also several strong peaks at around $10$, $14$ , $21$, and $24$  day$^{-1}$ ($f_7=9.9353, f_8=13.8850, f_{21}=21.3112, f_{24}=24.4334$ day$^{-1}$, respectively). Many low-amplitude peaks are present across the whole spectrum. Since there are significant frequencies close to $f_{\rm Nyquist}$, we also check the super-Nyquist spectrum (Figure 3). All the frequencies above $f_{\rm Nyquist}$ have lower amplitudes than their mirror-reflected counterparts. Thus the peaks contained in $< f_{\rm Nyquist}$ range are indeed real pulsations.

We identified the combination frequencies in the form of $f_k=mf_i \pm nf_j$ or $f_i \pm mf_{\rm orb}$, where $m,n$ are integers satisfying $1\le (m, n)\le 2$ (Table 4). These combination frequencies are highlighted in Fig 3, and they generally have low amplitudes. The majority of the combination frequencies are in the form of $f_i \pm mf_{\rm orb}$, and they are due to the amplitude modulation by eclipses. We did not find obvious coupling between p and g-modes shown as combination frequencies. However, if these combination frequencies were present, it would be a clear indication that the pulsations originate from the same star.

The overall appearance of the spectrum agrees with that of an evolved $\gamma$ Dor/$\delta$ Scuti hybrid star of $T_{\rm eff}\sim 7000$ K (e.g., KIC~9851944 in Guo et al.\ 2016). We next discuss the nature of these pulsations  and examine whether the observed frequencies agree with the unstable range of pulsation modes (Section 3.2).

\subsection{A General Overview of Unstable Modes}

The upper panel of Figure 4 shows the evolution of theoretical frequencies of $l=0,1,2$ modes (black circles, red triangles, and green squares, respectively) associated with a non-rotating interior model of $1.70M_{\odot}$. The calculation was performed from Zero Age Main Sequence (ZAMS)($R=1.5R_{\odot}$) to Terminal Age Main Sequence (TAMS) ($R=2.3R_{\odot}$). The interior models and the oscillation frequencies are calculated with MESA (v8118) (Paxton et al.\ 2011, 2013, 2015) and GYRE (v4.3) (Townsend \& Teitler 2013), respectively. We adopt the solar mixtures of Grevesse \& Sauval (1998), with the metal mass fraction $Z=0.02$ and an initial Helium abundance of $Y=0.28$. The OPAL opacity table (Iglesias \& Rogers 1996) is used, and convective overshooting is neglected. The mixing length parameter is fixed to the solar-calibrated value of $\alpha_{\rm MLT}=1.8$. The set-up above is consistent with observations of KIC~9592855 since we derive a solar metallicity\footnote{We should keep in mind that the observed photospheric abundance patterns do not necessarily reflect the global metallicity (e.g., chemically peculiar stars). However, we find that models with low metallicty ($Z=0.01$) have problems in exciting the observed p-modes.} and a mass of about $1.7M_{\odot}$ for both components. The filled symbols indicate unstable modes. Note that $\alpha_{\rm MLT}$ is very important in the modeling of mode excitations of $\delta$ Scuti stars. According to the calibration of Trampedach et al.\ (2014) through 3-D simulations, this parameter could vary from 1.6 to 2.0 depending on different effective temperatures and surface gravities. 
 
The stability parameter, $\eta$, is defined as the normalized growth-rate defined in Stellingwerf (1978) (the integration of differential work $\frac{dW}{dr}$ over the whole star, $\eta=\int^R_0(\frac{dW}{dr})dr/\int_0^R|\frac{dW}{dr}|dr$). Positive values of $\eta$ indicate modes are excited.  
We find that the $\eta$ values from GYRE's non-adiabatic calculation are not numerically well-behaved for modes of very low-frequencies\footnote{The collocation solvers in GYRE generally have better performance for non-adiabatic calculations than Magnus solvers.}. Thus we re-calculate the pulsation modes with Dziembowski's non-adiabatic code (Dziembowski 1971,1977) NADROT.
The theoretical frequencies from the two codes show excellent agreement, but the $\eta$ values from NADROT are more robust.

We choose the best-matching interior model for the secondary star with $M_2=1.7M_{\odot},R_2=1.96R_{\odot}$ from the above evolutionary sequence. The associated pulsation frequencies and the variations of $\eta$ are shown in the lower panel of Figure 4.
For convenience, we also over-plot the observed Fourier spectrum and scale it to have a maximum amplitude of 0.4.
We can roughly divide the variations of the instability parameter $\eta$ into two regimes. For p-modes, with frequencies $f \ge 14 $ day$^{-1}$, the variation of $\eta$ is independent of spherical degree $l$. Several low order p-modes ($p_1,p_2,p_3,p_4$) are excited. Note that the $p_5$ radial mode from NADROT is unstable, but it only has a very small positive $\eta$. In GYRE , it is barely damped. In the g-mode regime ($f < 14$ day$^{-1}$), the stability parameters $\eta$ are $l$-dependent, and this is especially true for high-order modes. Only the lowest-order g-modes are unstable ($n=-1,-2$), and all higher-order modes ($< 10$ day$^{-1}$) have negative $\eta$, i.e., they are damped. 

The theoretically unstable modes are low-order p- and g-modes, from $\sim 10$ day$^{-1}$ to $\sim 30$ day$^{-1}$. They can explain the observed frequency peaks from $\sim 10$ day$^{-1}$ to $\sim 24$ day$^{-1}$. The low-frequency modes observed ($f<10$ day$^{-1}$), especially the cluster of high-amplitude modes around 2 day$^{-1}$, cannot be explained by our calculations. However, there is a local peak of $\eta$ near this observed cluster of frequencies. It is not surprising that our non-adiabatic calculation based on the frozen-convection approximation cannot excite low-frequency g-modes. This is actually the major problem in our theory for the mode excitation mechanism: the  theory can well explain the excitation of pulsations in $\beta$ Cephei stars, Slowly Pulsating B-stars (SPB), and hot $\delta$ Scuti stars (e.g. Pamyatnykh 1999), but it cannot excite the low-frequency g-modes observed in $\gamma$ Dor stars and $\delta$ Scuti/$\gamma$ Dor hybrids. Guzik et al.\ (2000)
attributed the driving of $\gamma$ Dor stars to the `convective blocking mechanism', and a robust theory of time-dependent pulsation-convection interaction is still desirable, e.g., Dupret et al.\ (2005) and Xiong et al.\ (1997, 2015, 2016). The recent progress in the time-dependent convection theory emphasizes the importance of turbulent pressure in the mode excitation of cool $\delta$ Scuti and $\gamma$ Dor stars (Xiong et al. 2016). In particular for pressure modes, turbulent pressure affects the excitation of high-order modes (Antoci et al. 2014). 

With a mass and radius of $M_1=1.72 M_{\odot}$ and $R_1=1.53R_{\odot}$, the primary star is essentially on the ZAMS, and if it were pulsating, the unstable range of p-modes should be at about $35-40$ day$^{-1}$ ($p_3,p_4$). This range is somewhat higher than the frequencies observed ($0-25$ day$^{-1}$). Thus from the perspective of the theoretically unstable frequency range,  the observed pulsations of KIC~9592855 agree with the theoretical predictions for the secondary star, but we cannot completely rule out the possibility that the primary is pulsating. Determination of the pulsation origin is a common problem in the analysis of pulsating binaries. This issue can be resolved by examining the light travel time effect (phase modulation or amplitude modulation, Shibahashi \& Kurtz 2012; Murphy et al.\ 2014; Schmid et al.\ 2015). Regrettably, this method does not work for KIC 9592855 with such a short orbital period ($P=1.2$ d). The eclipse mapping method (Reed 2005; B{\'{\i}}r{\'o} 
\& Nuspl 2011), i.e., by inspecting the pulsation amplitude/phase variations in eclipses, can in principle determine the origin of pulsations and even identify the pulsation modes. But it is technically challenging and still awaits its application to a real star.  Note that intrinsic amplitude and phase variations can also be present in $\delta$ Scuti stars and compactor pulsators (white dwarf and hot subdwarf stars) as recently studied in Bowman et al.\ (2016) and Zong et al.\ (2016a, 2016b). 

The above comparison between the observed and theoretical frequency range of unstable modes are only approximate (as well as in Fig. 7 later). If we include rotation, the growth rate $\eta$ of splitted mode ($m=(+1, -1; +2,+1,-1,-2)$) can be calculated from the perturbed eigenfunctions, and the results are almost the same as the central axisymmetric (m=0) modes. The excited frequency range is expected to be wider by $\sim 2\times (1-C_{\rm nl})f_{\rm rot}\approx 2f_{\rm rot}=1.6$ day$^{-1}$ for p-modes. Thus this small correction will not change our conclusion above.

One caveat: as will be detailed in Section 4, this system may have gone through the binary evolution with mass transfer. The above analysis is based on the assumption that an interior structure model from single star evolution with the observed mass and radius can be used to infer the pulsational spectrum. It has been shown that this assumption is probably valid, as the p-mode pulsations of a post-mass transfer, rejuvenated $\delta$ Scuti star in KIC~8262223 (Guo et al.\ 2017) can indeed be explained by using a model from single star evolution.  However, we could think of two examples in which single-star-evolution models cannot be used. Firstly, the mass gainer could accrete some Helium-abundant material from its companion which will affect the mode excitation through the $\kappa$-mechanism. Secondly, since period spacings depend on the g-mode cavity and thus on the extent of convective core. In some cases, the mass gainer may have a larger convective core than its single-star-evolution counterpart after mass transfer. This will change the vertical level of period spacings.
The delicate effects of mass-transfer on the pulsations need further investigation. We also leave the detailed asteroseismic modeling of the observed individual frequencies to a future study and concentrate on the equally-spaced g-mode pattern in Section 3.3.

\subsection{Identification of Prograde Dipole g-modes Affected by Rotation}

In the g-mode regime, the dominant frequency peaks are around $2$ day$^{-1}$ (Figure 3). These peaks are almost equally spaced in period. In Figure 5, we have marked these peaks with red dotted lines. In the order of increasing period, they are $f_{24}=30.538,f_{9}=32.3392,f_{*}$\footnote{This frequency peak has a signal-to-noise ratio slightly lower than 4 ($S/N\sim 3.6$) with our adopted noise level. Since it complies with the period spacing pattern, it is very likely to be a real pulsation. We thus include it but use a different notation $f_{*}$.}$=34.0459,f_{3}=35.6630, f_{6}=37.2075, f_{1}=38.6991, f_{18}=40.1458$ in units of $1000$ seconds. Table 3 contains the details of these pulsations. A careful examination reveals that the period spacings decrease monotonically, from $dP=1.8$ at period of $\sim 3200$s  to $dP=1.6$ at period $\sim 4000$s. In the period-spacing vs. period diagram ($dP-P$) shown in Figure 6, this decline of period spacings manifests itself as a downward (negative) slope. This is clearly the signature of rotational effects on high-order g-modes. A linear fit to the observed $dP-P$ yields a slope of $\Sigma=-0.043$. This slope directly relates to the internal rotation of the star as will be elaborated below. 

With the goal of modeling the slope of the period spacing pattern, and largely following Van Reeth et al.\ (2016), we begin with the asymptotic relation for high-order g-modes (Tassoul 1980) without rotation. The pulsation period is given by: 
 \begin{equation}
P_{nl} \approx \frac{\Pi_0}{\sqrt{l(l+1)}}(n+0.5),
\end{equation}
where $n$ is the radial order, $l$ is the spherical degree$, \Pi_0=2\pi^2(\int_{r_1}^{r_2}\frac{N}{r}dr)^{-1}$, and $N$ is the Brunt-V\"ais\"al\"a frequency. The integration in the expression of $\Pi_0$ is performed within the g-mode propagation cavity $r_1 \rightarrow r_2$.  The asymptotic period spacing is:
 \begin{equation}
\Delta\Pi_l=\frac{\Pi_0}{\sqrt{l(l+1)}}.
\end{equation}
The observed period spacing is usually indicated by $dP$, but the theoretical asymptotic period spacing is denoted by $\Delta\Pi_l$.
The period spacing $\Delta\Pi_l$ is a constant for fixed spherical degree $l$ and for a fixed stellar interior model. For $\gamma$ Dor stars, which generally have a convective core, $\Delta\Pi_l$ decreases monotonically from ZAMS to TAMS. This is mainly because the g-mode cavity becomes deeper and larger as a result of the retreat of the convective core,  making the integration $\int \frac{N}{r} dr$ larger and thus $\Pi_0$ smaller (e.g., Fig.10 in Schmid et al.\ 2016).

With rotation, the period spacing pattern in the observer's (inertial) frame contains a linear trend, and the slope of this trend is either negative for zonal ($m=0$) and prograde\footnote{We assume the mode eigenfunctions have the time dependence of $\propto e^{-i\omega t}$.} ($m>0$) modes or mainly positive for retrograde modes ($m<0$).
The observed slope results from the transformation of pulsation frequencies from the co-rotating frame ($co$) to the inertial frame ($in$): $f_{in}=f_{co}+mf_{rot}$,
and also from the reduction or increase of pulsation periods due to rotation. The latter can be characterized by the relation which invokes the traditional approximation:

\begin{equation}
P_{co}\approx \frac{\Pi_0}{\sqrt{l_{\rm eff}(l_{\rm eff}+1)}}(n+0.5),
\end{equation}
where $l_{\rm eff}=(\sqrt{1+4\lambda_{l,m,s}}-1)/2$ is the effective spherical degree introduced by Townsend (2005). With the introduction of $l_{\rm eff}$, the asymptotic relation for period is in the same form of the non-rotating case of eq.\ (1). $s=2f_{rot}/f_{co}$ is the spin parameter, and 
$\lambda_{l,m,s}$ are the eigenvalues of the Laplace tidal equations, which are the reciprocal of the eigenvalues of matrix $\bm{W}^{-1}$ defined in, e.g., Unno et al.\ (1989) (eq. 34.29, 34.30).  In the zero-rotation limit ($s\rightarrow 0$), $\lambda_{l,m,s}\rightarrow l(l+1)$ and $l_{\rm eff} \rightarrow  l$.  For prograde and zonal g-modes, $\lambda_{l,m,s}$ increases with $s$ except for prograde sectoral  modes $(l=m)$.

To model the observed $dP-P$, we generate synthetic g-modes with a grid of consecutive radial orders $n$ in the co-rotating frame (eq. 3) and transform them to the inertial frame. This grid of periods can then be interpolated to the observed values of pulsation periods, and a $\chi^2$-minimization can be performed with $P$ and $dP$ as independent and dependent variables, respectively. This modeling procedure only involves two free parameters: the rotation frequency $f_{\rm rot}$ and the asymptotic period spacing $\Delta \Pi_{l}$ (or  $\Delta \Pi_{0}$). It is also assumed that $l$ and $m$ are known. In practice, although multiple combinations of $l$ and $m$ may fit the observed $dP-P$ equally well, we can usually determine the correct $(l,m)$ combination by comparing the resulting $\Delta\Pi_{l}$ to the expected theoretical values. 
This method does not need the time-consuming calculation of evolutionary interior models and thus can be applied to a large sample of stars. For example, Van Reeth et al.\ (2016) did an ensemble analysis of internal rotations by using the period spacing pattern for a sample of 68 $\gamma$ Dor stars.

Ouazzani et al.\ (2017) did extensive calculations of pulsation periods by using both the traditional approximation and their 2-D code. Their results can give some guidance on modeling KIC~9592855. First, there is a near one-to-one relation between the slope $\Sigma$ and rotation frequency $f_{\rm rot}$, and this relation is relatively insensitive to metallicity and additional mixing near the edge of the convective core (convective overshooting, microscopic diffusion, etc.).
Second, only zonal and prograde modes have a downward slope, i.e., negative $\Sigma$. The slope of zonal modes never exceeds $-0.025$; thus the observed $\Sigma=-0.043$ in KIC~9592855 indicates prograde modes.

From the perspective of geometric cancellation over the stellar disk, usually only $l=1$ and $l=2$ g-modes\footnote{We cannot exclude the presence of higher $l$ values as these do not suffer of total cancellation.}  can be observed in the broad-band photometric data of {\it Kepler}.  KIC~9592855 also has a high inclination ($i=73\fdg 3$), assuming pulsation-spin-orbit alignment, which indicates that ($l=1,m=1$) modes have a higher visibility than ($l=1,m=0$) modes (Gizon \& Solanki 2003). Similarly, ($l=2,m=2$) modes are more visible than ($l=2,m=0$) and ($l=2,m=1$) modes.

Thus assuming $l=1,m=1$, we fit the observed $dP-P$ by using the procedure mentioned above. This results in $f_{\rm rot}=0.8\pm 0.1$ day$^{-1}$ and $\Delta\Pi_{l=1}=3.0\pm 0.2$ (in 1000 sec). The best-fitting model and the contour plot of $\chi^2$ values are shown in the right two panels of Figure 6. The left panel shows the period \'{e}chelle diagram, with periods modulo a fixed spacing of 1600 seconds. The g-modes belonging to the consecutive prograde dipole modes form a parabolic ridge, indicating period spacings that decrease linearly with periods. This is the signature of fast rotation in the \'{e}chelle diagram. For more such examples, please refer to Bedding et al.\ (2015).

We find that ($l=2,m=2$) modes can also fit the $dP-P$ pattern with similar chi-square values, but the resulting asymptotic period spacing is $\Delta\Pi_{l=2}=3200$ seconds, which is too large to be associated with $l=2$. We can thus safely discard this possibility. This is the same method used in Van Reeth et al.\ (2016) for mode differentiation.
The identification of $l=1,m=1$ g-modes in KIC~9592855 also agrees with their discovery that the prograde dipole modes prevail in the $\gamma$ Dor stars.


Note that the method discussed above is only approximate. By comparing with theoretical frequencies from evolutionary models, detailed modeling of individual g-mode frequencies can refine the rotation rates and asymptotic period spacings.
The $f_{\rm rot}$ measured from the period spacing slope of g-modes is an averaged value over the Brunt-V\"ais\"al\"a frequency $N$. It can thus be treated as the rotation rate of the convective core boundary, since this transition region with its chemical composition gradient has the largest contribution to $N$ and thus to the period spacing (Miglio et al.\ 2008). It is expected that within the chemically homogeneous radiative envelope the rotation rate should be nearly uniform (Pamyatnykh et al.\ 2004). A large rotational gradient may occur mainly in the transition zone around the convective core. 

It is interesting to note that the derived interior rotation rate $f_{\rm rot}=0.8\pm 0.1$ day$^{-1}$ is very close to the surface rotation frequency, which is the same as the orbital frequency $f_{\rm orb}=0.8201$ day$^{-1}$. A comparison with Fig 9 in Ouazzani et al.\ (2017) yields a slightly higher rotation rate of $0.9-1.0$ day$^{-1}$. Thus the core has the same or slightly faster rotation rate than the envelope. This is the first measurement of the internal rotation of F- or A-type stars in a synchronized close binary system. The two F-star components in the eccentric binary system KIC~10080943 (Schmid et al.\ 2015) also show near-uniform interior rotation, as revealed from the splitting of g-modes.

For single stars, near uniform rotation has been measured in a few F- or A-type stars (KIC~9244992 in Saio et al.\ 2015; KIC~1145123 in Kurtz et al.\ 2015; KIC~7661054 in Murphy et al.\ 2016) as well as a B-type star HD 157056 (Briquet et al.\ 2007). These measurements are mostly from rotational splittings of envelope-sensitive p-modes and core-sensitive g-modes, which limit the application to slow rotators. Several case studies of B-type stars reveal faster rotating cores (HD 129929 in Dupret et al.\ 2004; $\nu$ Eridani in Pamyatnykh et al.\ 2004).
On the contrary, Triana et al.\ (2015) discovered that the SPB star KIC~10526294 has a much faster-rotating envelope that its core ($\sim$ three times) and in the opposite direction. Recently, P{\'a}pics et al.\ (2017) reported a sample of SPB stars with period spacing patterns affected by rotation. However, the $dP-P$ patterns are not fully exploited to derive core-to-envelope differential rotation. Ouazzani et al. (2017) derived the internal rotation rates for four $\gamma$ Dor stars (KIC 4253413, KIC 6762992, KIC 5476299, and KIC 4177905), but no information on the envelope rotation is provided. For a recent review, please refer to Aerts et al. (2017).

Massive stars possess a convective core and radiative envelope, and intermediate mass stars have the same structure but also with a shallow convective envelope near the surface. The numerical simulations by Rogers et al.\ (2013, 2015) prove that the transport of angular momentum by internal gravity waves (IGW) can explain all the aforementioned rotation profiles in single stars. The angular momentum transport in early-type binary stars involves the consideration of tidally excited gravity waves instead of the convection-driven IGW in the single-star case. These tidally driven oscillations are excited near the convective-radiative interface and propagate inside the radiative envelope and dissipate due to radiative diffusion and non-linearity near the surface (Goldreich \& Nicholson 1989). They are the primary source for the orbit and spin decay and have been used to explain successfully the circularization and synchronization of early-type binaries (Zahn 1975, 2008)

Goldreich \& Nicholson (1989) suggest that the stellar surface may be synchronized first, and thus the stars spin down or up from outside inward. In line with this argument, Kallinger et al.\ (2017) found that the surface layer of the SPB star in the binary system HD 201433 has been spun up by its companion while the inner layers have a much slower and near-rigid rotation. The spin history of KIC~9592855 may be very complex, as it may involve binary evolution with mass transfer as will be discussed in the next section. Possibly, the secondary star in KIC~9592855 had a higher primordial rotation rate, and it has been spun down from the surface inward by its companion and synchronization is achieved before the mass transfer begins. Currently, KIC~9592855 has finished the mass transfer, and thus the secondary has a uniform rotation rate. The measured surface and internal rotation rates can give us a hint about the complicated binary evolutionary history.



 


In circular and synchronized binaries, the static equilibrium tide mainly affects the high-frequency p-modes, and it can be treated as a perturbation to the second order of rotation (rotational distortion and tidal distortion; Saio 1981). The effect of distortion on g-modes is small. For binaries with very eccentric orbits, the dynamical tide can induce oscillation modes of mainly $l=2$ (Welsh et al.\ 2011; Hambleton et al.\ 2016; Guo et al.\ 2017; Pablo et al.\ 2017). KIC~9592855 apparently falls into the former regime.

\subsection{Evolution}

To determine the evolutionary state of KIC~9592855, we show the evolution of stellar radius and the asymptotic period spacing of $l=1$ modes $\Delta\Pi_{l=1}$ in the lower panel of Figure 7. 
The two sets of tracks can be distinguished by their corresponding colors, where red tracks are for the radius evolution and black tracks are for the period spacing evolution. The solid, dotted, and dashed tracks have associated stellar masses of 1.8$M_{\odot}$, 1.7$M_{\odot}$, and 1.6 $M_{\odot}$, respectively. The observed radii of the primary and secondary star, with $\pm 1 \sigma$ credible regions, are indicated by the green and blue shaded horizontal bars, respectively.

We can use a set of vertical lines to infer the age of this system (if they were from single star evolution), taking into account the observed radii of the two stars.
The observed radius of the primary star indicates it is close to ZAMS, with an age less than $2 \times 10^8$ yrs, and this can be seen from the deep green region where the evolutionary tracks of stellar radii intersect the green bar. However, to explain the observed radius of the secondary star ($R_2=1.96R_{\odot}$, the blue bar) using the red evolutionary tracks,  we have to adopt a much older age, about 0.7 to 1.15 Gyr (deep blue region). The two stars have an almost identical mass and very different radii, and cannot be explained by a pair of coeval models from single star evolution.
With an orbital period of only 1.2 days, and an orbital separation of $\sim 7$ solar radii, one natural explanation of this contradiction is the binary star evolution with mass transfer. This binary seems to be close to the state of minimum orbital period/separation in the binary evolution when the mass ratio approaches unity. We also repeat the above steps using evolutionary tracks with convective core overshooting ($f_{ov}=0.02$). Since the main-sequence is extended in this case, the age discrepancy of the two stars actually becomes larger. Our efforts of resolving this discrepancy by including an equatorial rotational velocity of $80$ km s$^{-1}$ or a sub-solar metallicity (Z=0.01) in the MESA models bear no fruit.

Note that we also show the observed period spacing of dipolar modes ($\Delta\Pi_{l=1}=3.0 \pm 0.2) \times 1000$ seconds, the gray horizontal bar). All models with ages less than $2 \times 10^8$ yrs in the deep green region have period spacings values ($3.4-3.6$) higher than the observation. The gray bar intersects the black tracks (theoretical $\Delta\Pi_{l=1}$ with different masses) at ages of about $7.5 \times 10^8$ to $1.4 \times 10^9$ yrs, as indicated by the deep gray region. This age range overlaps the previous age range of the secondary star from comparison of stellar radius (deep blue region). It means that evolutionary models with age from $7.5 \times 10^8$ to $1.15 \times 10^9$ yrs can explain both the observed radius and dipolar mode period spacing. This age range is very similar to the age range when the observed p-mode frequencies (solid and dotted horizontal lines) approximately match the predicted unstable range (orange shaded region) as shown in the upper panel of Fig. 7 (from about $8.0 \times 10^8$ to $1.2 \times 10^9$ yrs). The consistency of stellar radius, period spacing, and unstable frequency range strongly suggests that most, if not all, of the observed g- and p-modes are from the secondary star.

The ages inferred above cannot be adopted as the age of this binary system, but the age discrepancy discussed above reveals convincingly the mass transfer history of this binary. The primary star has typical parameters ($M=1.72M_{\odot},R=1.53R_{\odot}$) of a ZAMS star. It is thus likely that it has been rejuvenated in the past, i.e., it is formed by accreting mass from the secondary star.
The secondary has a large filling factor ($f_2=0.565$), and it may have finished the mass transfer in the not-too-distant past. The system thus becomes detached, and its orbital period will lengthen from the current value. In a sense, KIC~9592855 roughly resembles KIC~8262223 (Guo et al.\ 2017), another post-mass transfer $\delta$ Scuti pulsator with comparable orbital period and separation. The two binaries may have similar evolutionary histories.

\section{Conclusions and Prospects}

We performed photometric and spectroscopic analysis of the post-mass transfer $\gamma$ Dor/$\delta$ Scuti hybrid pulsating eclipsing binary KIC~9592855. The pulsation spectrum after subtracting the binary light curve shows strong low-frequency g-modes and low-order p-modes. We identify these dominant g-modes as prograde dipole modes. Their near-equally spaced period pattern, which likely arises from the secondary star, reveals a near-uniformly rotating core and envelope. The derived asymptotic period spacing of dipole modes also supports the argument that the  observed g-modes are from the secondary. The identical mass but very different radii of the two components strongly suggest that this close binary has a mass-transfer history.

Much work remains to be done to understand this intriguing binary. Albeit challenging, it is worthwhile  to model observed p-modes. With a projected equatorial rotational velocity of $\sim 80$ km s$^{-1}$, perturbation to the second order is needed to account for the centrifugal distortion and tidal distortion. It is also highly desirable to perform a detailed binary evolutionary modeling similar to those done by St{\c e}pie{\'n} et al.\ (2017). If theoretical pulsation frequencies of those observed g-modes or even p-modes can be calculated at different stages of mass transfer, our theory of the close binary evolution can be confronted with observations and thus refined. 
In this work, only seven radial velocities have been measured. More spectroscopic observations covering the full orbit can improve the fundamental parameters of this binary. The light curve of KIC~9592855 contains relatively wider eclipses, and this makes it a good candidate to implement the eclipse mapping technique.

 
\acknowledgments 
 
We thank the referee for helpful suggestions and comments. We thank the {\it Kepler} team for creating the excellent photometric data used here. We also express thanks to Jerry Orosz for sharing his ELC code; to Gerald Handler, Alexey Pamyatnykh and Wojciech Dziembowski for providing the NADROT code and helpful discussions; to Jakub Ostrowski  and Wojciech Szewczuk for sharing their nice Fortran program connecting MESA and NADROT; to Bill Paxton, Rich Townsend et al.\ for making MESA and GYRE available; to 	Hideyuki Saio for helpful discussions on traditional approximation. Z.G. thanks Stephen Williams for help in using the ELC code. This work has 
been supported by the Polish NCN grant 2015/18/A/ST9/00578. This material is based upon work supported by the U.S.\
National Science Foundation under Grant No.~AST-1411654.
 
\clearpage

\begin{deluxetable}{lccccc} 
\tabletypesize{\small} 
\tablewidth{0pc} 
\tablenum{1} 
\tablecaption{Atmospheric Parameters\label{tab1}} 
\tablehead{ 
\colhead{Parameter}   & 
\colhead{Primary\tablenotemark{b} }     &
\colhead{Secondary\tablenotemark{b} }     &
\colhead{Primary\tablenotemark{c}}      &
\colhead{Secondary\tablenotemark{c}}      &

}
\startdata 
$T_{\rm eff}$ (K)               \dotfill & $7567 \pm 65$ & $7037 \pm 55$  & $7604 \pm 73$ & $7047 \pm 64$\\ 

$\log g$ (cgs)  \dotfill & $4.3\tablenotemark{a}$  & $4.1\tablenotemark{a}$ &$4.3\tablenotemark{a}$  & $4.1\tablenotemark{a}$       \\ 

$v \sin i$ (km s$^{-1}$)     \dotfill & $61 \pm 10$        & $62 \pm 10$ & $62 \pm 11$ & $61 \pm 10$          \\ 

$\rm[Fe/H]$ \dotfill & $-0.05 \pm 0.04$        & $ -0.05 \pm 0.04$ & $ -0.04 \pm 0.03$   & $ -0.03 \pm 0.02$               \\   
\enddata 
\tablenotetext{a}{\ Fixed.}
\tablenotetext{b}{\ Genetic algorithm.}
\tablenotetext{c}{ \ Levenberg-Marquardt algorithm.}
\end{deluxetable} 


\begin{deluxetable}{lccc} 
\tabletypesize{\small} 
\tablewidth{0pc} 
\tablenum{2} 
\tablecaption{Binary Model Parameters \label{tab2}} 
\tablehead{ 
\colhead{Parameter}   & 
\colhead{Primary (1)}      &
 \colhead{Secondary (2)}  &
\colhead{System}       
}
\startdata 
Orbital Period (days) & & &$1.21932483 \tablenotemark{a}\pm 0.00000002$\\
Time of primary minimum, $T_0$ (BJD-2400000) & & &$55691.664073 \tablenotemark{a}\pm 0.00007$   \\
Orbital eccentricity, $e$              & & &$0.0$\tablenotemark{a}     \\
Orbital inclination (degree), $i$ & & & $73.25\tablenotemark{b}\pm 0.07$\\
Semi-major axis ($R_\odot$), $a$ & & & $7.24\pm 0.09$\\
Mass ratio $q=M_2/M_1=K_1/K_2$               & & &$0.99\tablenotemark{b} \pm 0.03$    \\ 
Systemic velocity (km $s^{-1}$), $\gamma $ & & &$8.7\tablenotemark{b}\pm 0.6$\\
Mass ($M_\odot$)              & $1.72 \pm 0.07$             & $1.71 \pm 0.06$    \\ 
Radius ($R_\odot$)               & $1.53\pm 0.03$       & $1.96 \pm 0.03$    \\

Filling factor, $f$     &$0.426\tablenotemark{b}\pm 0.006$    &$0.565\tablenotemark{b}\pm 0.004$\\
Gravity brightening coefficient, $\beta$     &$0.08\tablenotemark{a}$    &$0.08\tablenotemark{a}$\\

Bolometric albedo, $l$ & $0.32\tablenotemark{b}\pm 0.04$   & $0.34\tablenotemark{b}\pm 0.05$    \\

$T_{\rm eff}$ (K)                    & $7300\tablenotemark{a} $ & $7202\tablenotemark{b} \pm 70$\\
$\log g$ (cgs)  & $4.30\pm 0.03$     & $4.09\pm 0.03$     \\ 
Synchronous $v \sin i$ (km s$^{-1}$)   &   $61 \pm 1$   &   $78\pm 1$              \\ 
Velocity semi-amplitude $K$ (km~s$^{-1})$  &$143.6\tablenotemark{b}\pm 2.4 $ & $144.3\pm 4.3$\\ 
\enddata 
\tablenotetext{a}{Fixed parameters: $P,T_0,e,\beta_1,\beta_2,T_{\rm eff1}$.}
\tablenotetext{b}{Free parameters: $i,f_1,f_2,l_1,l_2,T_{\rm eff2},K_1,q,\gamma$.}
\end{deluxetable}  

\clearpage

\begin{deluxetable}{lcccccccc} 
\rotate
\tabletypesize{\small} 
\tablewidth{0pc} 
\tablenum{3} 
\tablecaption{Near Equally-spaced Prograde g-modes\label{tab3}} 
\tablehead{ 
\colhead{} &
\colhead{Period (day)}   & 
\colhead{Period ($1000$ sec)}      &
\colhead{Frequency (d$^{-1}$)}      &
\colhead{Amplitude ($\Delta F/F$)($10^{-3}$)}      &
\colhead{Phase (rad/$2\pi$)} & 
\colhead{S/N} &
}
\startdata 
$f_{      24}$    &$0.353450(3)$ & $30.53811(25)$ & $2.82925(2)$ & $0.093(16)$ & $0.480(78)$ & $10.2$ & \\
$f_{       9}$    &$0.374296(2)$ & $32.33915(14)$ & $2.67168(1)$ & $0.192(16)$ & $0.422(39)$ & $20.2$ &  \\
$f_{      *}$   &$0.394050(10)$  & $34.04593(88)$ & $2.53775(7)$ & $0.035(17)$ & $0.381(224)$ & $3.6$ &  \\
$f_{       3}$ &$0.412766(1)$    & $35.66299(6)$ & $2.42268(1)$ & $0.609(17)$ & $0.190(13)$ & $60.9$ &  \\
$f_{       6}$  &$0.430642(1)$   & $37.20746(11)$ & $2.32212(1)$ & $0.353(18)$ & $0.419(23)$ & $34.5$ &  \\
$f_{       1}$  &$0.447906(1)$  & $38.69912(3)$ & $2.23261(1)$ & $1.522(18)$ & $0.283(5)$ & $145.7$ &  \\
$f_{      18}$ &$0.464650(4)$    & $40.14575 (37)$ & $2.15216(2)$ & $0.125(18)$ & $0.479(68)$ & $11.8$ & \\

\enddata 
\tablenotetext{a}{Note: $f_{*}$ has S/N $<4$ but is included because of its relation to the other g-mode frequencies (see Section 3.3)}
\end{deluxetable} 

\clearpage

\begin{deluxetable}{lccccccc} 
\tabletypesize{\small} 
\tablewidth{0pc} 
\tablenum{4} 
\tablecaption{Other Significant Oscillation Frequencies\label{tab4}} 
\tablehead{ 
\colhead{}   & 
\colhead{Frequency (d$^{-1}$)}      &
\colhead{Amplitude ($\Delta F/F) (10^{-3}$)}      &
\colhead{Phase (rad/$2\pi$)} & 
\colhead{S/N} &
\colhead{Comment} &
}
\startdata 
$f_{       1}$    & $2.23261 \pm 0.00001$ & $1.522 \pm 0.018$ & $0.283\pm 0.005$ & $145.7$ & $l=1,m=1$ \\
$f_{       2}$    & $24.43343 \pm 0.00001$ & $0.290 \pm 0.008$ & $0.353\pm 0.012$ & $65.0$ & $ $ \\
$f_{       3}$    & $2.42268 \pm 0.00000$ & $0.609 \pm 0.017$ & $0.190\pm 0.013$ & $60.9$ & $l=1,m=1$ \\
$f_{       4}$    & $21.31124 \pm 0.00001$ & $0.240 \pm 0.008$ & $0.473\pm 0.015$ & $52.2$ & $ $ \\
$f_{       5}$    & $1.95288 \pm 0.00001$ & $0.452 \pm 0.019$ & $0.502\pm 0.020$ & $40.6$ & $ $ \\
$f_{       6}$    & $2.32212 \pm 0.00001$ & $0.353 \pm 0.018$ & $0.419\pm 0.023$ & $34.5$ & $l=1,m=1$ \\
$f_{       7}$    & $9.93593 \pm 0.00001$ & $0.308 \pm 0.019$ & $0.804\pm 0.029$ & $27.7$ & $ $ \\
$f_{       8}$    & $13.88504 \pm 0.00001$ & $0.218 \pm 0.015$ & $0.865\pm 0.032$ & $25.2$ & $ $ \\
$f_{       9}$    & $2.67168 \pm 0.00001$ & $0.192 \pm 0.016$ & $0.422\pm 0.039$ & $20.2$ & $l=1,m=1$ \\
$f_{      10}$    & $24.43296 \pm 0.00001$ & $0.085 \pm 0.008$ & $0.075\pm 0.042$ & $18.9$ & $ $ \\
$f_{      11}$    & $6.56103 \pm 0.00001$ & $0.108 \pm 0.011$ & $0.201\pm 0.046$ & $17.4$ & $ $ \\
$f_{      12}$    & $4.92074 \pm 0.00002$ & $0.096 \pm 0.011$ & $0.283\pm 0.051$ & $15.5$ & $f_{11}-2f_{orb}$ \\
$f_{      13}$    & $1.28384 \pm 0.00002$ & $0.204 \pm 0.023$ & $0.708\pm 0.052$ & $15.5$ & $ $ \\
$f_{      14}$    & $19.18962 \pm 0.00002$ & $0.087 \pm 0.010$ & $0.590\pm 0.053$ & $15.0$ & $ $ \\
$f_{      15}$    & $3.93411 \pm 0.00002$ & $0.102 \pm 0.012$ & $0.894\pm 0.056$ & $14.1$ & $ $ \\
$f_{      16}$    & $18.16594 \pm 0.00002$ & $0.103 \pm 0.013$ & $0.887\pm 0.057$ & $13.9$ & $ $ \\
$f_{      17}$    & $0.74094 \pm 0.00002$ & $0.193 \pm 0.026$ & $0.811\pm 0.064$ & $12.5$ & $ $ \\
$f_{      18}$    & $2.15216 \pm 0.00002$ & $0.125 \pm 0.018$ & $0.479\pm 0.068$ & $11.8$ & $l=1,m=1$ \\
$f_{      19}$    & $10.58986 \pm 0.00002$ & $0.128 \pm 0.020$ & $0.958\pm 0.071$ & $11.2$ & $ $ \\
$f_{      20}$    & $24.43401 \pm 0.00002$ & $0.049 \pm 0.008$ & $0.799\pm 0.073$ & $10.9$ & $ $ \\
$f_{      21}$    & $20.82988 \pm 0.00002$ & $0.052 \pm 0.008$ & $0.367\pm 0.073$ & $10.9$ & $f_{14}+2f_{orb}$ \\
$f_{      22}$    & $18.81166 \pm 0.00002$ & $0.068 \pm 0.011$ & $0.111\pm 0.074$ & $10.7$ & $ $ \\
$f_{      23}$    & $8.20127 \pm 0.00002$ & $0.091 \pm 0.015$ & $0.506\pm 0.076$ & $10.6$ & $f_{11}+2f_{orb}$ \\
$f_{      24}$    & $2.82925 \pm 0.00002$ & $0.093 \pm 0.016$ & $0.480\pm 0.078$ & $10.2$ & $l=1,m=1$ \\
$f_{      25}$    & $23.37736 \pm 0.00003$ & $0.040 \pm 0.007$ & $0.921\pm 0.086$ & $9.3$ & $ $ \\
$f_{      26}$    & $16.79271 \pm 0.00003$ & $0.074 \pm 0.014$ & $0.748\pm 0.090$ & $8.8$ & $ $ \\
$f_{      27}$    & $0.78244 \pm 0.00003$ & $0.125 \pm 0.026$ & $0.443\pm 0.097$ & $8.2$ & $f_3-2f_{orb}$ \\
$f_{      28}$    & $1.25028 \pm 0.00003$ & $0.109 \pm 0.023$ & $0.109\pm 0.097$ & $8.2$ & $ $ \\
$f_{      29}$    & $10.75605 \pm 0.00003$ & $0.088 \pm 0.020$ & $0.139\pm 0.103$ & $7.7$ & $f_7+f_{orb}$ \\
$f_{      30}$    & $11.65240 \pm 0.00003$ & $0.084 \pm 0.019$ & $0.015\pm 0.106$ & $7.5$ & $f_1-f_8$ \\
$f_{      31}$    & $11.18603 \pm 0.00003$ & $0.080 \pm 0.019$ & $0.675\pm 0.114$ & $7.0$ & $ $ \\
$f_{      32}$    & $0.01079 \pm 0.00003$ & $0.129 \pm 0.032$ & $0.368\pm 0.117$ & $6.8$ & $ $ \\
$f_{      33}$    & $8.76189 \pm 0.00003$ & $0.065 \pm 0.016$ & $0.780\pm 0.117$ & $6.8$ & $ $ \\
$f_{      34}$    & $3.28051 \pm 0.00003$ & $0.056 \pm 0.014$ & $0.876\pm 0.118$ & $6.8$ & $f_{12}-2f_{orb}$ \\
$f_{      35}$    & $0.69291 \pm 0.00004$ & $0.103 \pm 0.027$ & $0.930\pm 0.121$ & $6.6$ & $ $ \\
$f_{      36}$    & $14.76231 \pm 0.00004$ & $0.050 \pm 0.013$ & $0.724\pm 0.121$ & $6.6$ & $ $ \\
$f_{      37}$    & $21.73707 \pm 0.00004$ & $0.029 \pm 0.008$ & $0.975\pm 0.125$ & $6.4$ & $f_{25}-2f_{orb}$ \\
$f_{      38}$    & $0.59236 \pm 0.00004$ & $0.102 \pm 0.028$ & $0.978\pm 0.126$ & $6.3$ & $f_1-2f_{orb}$ \\
$f_{      39}$    & $7.70329 \pm 0.00004$ & $0.048 \pm 0.013$ & $0.117\pm 0.127$ & $6.3$ & $f_7-f_1$ \\
$f_{      40}$    & $13.88551 \pm 0.00004$ & $0.054 \pm 0.015$ & $0.605\pm 0.128$ & $6.2$ & $ $ \\
$f_{      41}$    & $20.49110 \pm 0.00004$ & $0.030 \pm 0.008$ & $0.515\pm 0.129$ & $6.2$ & $f_4-f_{orb}$ \\
$f_{      42}$    & $21.32186 \pm 0.00004$ & $0.028 \pm 0.008$ & $0.626\pm 0.131$ & $6.1$ & $ $ \\
$f_{      43}$    & $13.10256 \pm 0.00004$ & $0.059 \pm 0.017$ & $0.970\pm 0.132$ & $6.0$ & $ $ \\
$f_{      44}$    & $7.26424 \pm 0.00004$ & $0.042 \pm 0.012$ & $0.669\pm 0.134$ & $6.0$ & $f_7-f_9$ \\
$f_{      45}$    & $10.96873 \pm 0.00004$ & $0.066 \pm 0.020$ & $0.479\pm 0.137$ & $5.8$ & $ $ \\
$f_{      46}$    & $9.35431 \pm 0.00004$ & $0.061 \pm 0.018$ & $0.498\pm 0.137$ & $5.8$ & $ $ \\
$f_{      47}$    & $1.72521 \pm 0.00004$ & $0.067 \pm 0.020$ & $0.224\pm 0.139$ & $5.7$ & $ $ \\
$f_{      48}$    & $1.60257 \pm 0.00004$ & $0.068 \pm 0.021$ & $0.424\pm 0.143$ & $5.6$ & $f_3-f_{orb}$ \\
$f_{      49}$    & $1.41253 \pm 0.00004$ & $0.070 \pm 0.022$ & $0.512\pm 0.146$ & $5.5$ & $f_1-f_{orb}$ \\
$f_{      50}$    & $0.00558 \pm 0.00004$ & $0.102 \pm 0.032$ & $0.396\pm 0.147$ & $5.4$ & $ $ \\
$f_{      51}$    & $3.07663 \pm 0.00004$ & $0.046 \pm 0.015$ & $0.216\pm 0.151$ & $5.3$ & $ $ \\
$f_{      52}$    & $21.31071 \pm 0.00004$ & $0.024 \pm 0.008$ & $0.189\pm 0.153$ & $5.2$ & $f_{42}-2f_{50}$ \\
$f_{      53}$    & $11.57612 \pm 0.00005$ & $0.058 \pm 0.019$ & $0.101\pm 0.155$ & $5.2$ & $ $ \\
$f_{      54}$    & $2.49515 \pm 0.00005$ & $0.051 \pm 0.017$ & $0.869\pm 0.155$ & $5.1$ & $ $ \\
$f_{      55}$    & $12.26919 \pm 0.00005$ & $0.055 \pm 0.018$ & $0.558\pm 0.156$ & $5.1$ & $ $ \\
$f_{      56}$    & $1.24949 \pm 0.00005$ & $0.068 \pm 0.023$ & $0.420\pm 0.157$ & $5.1$ & $ $ \\
$f_{      57}$    & $2.38117 \pm 0.00005$ & $0.051 \pm 0.017$ & $0.225\pm 0.157$ & $5.1$ & $f_{17}+2f_{orb}$ \\
$f_{      58}$    & $21.05471 \pm 0.00005$ & $0.024 \pm 0.008$ & $0.431\pm 0.159$ & $5.0$ & $ $ \\
$f_{      59}$    & $9.96933 \pm 0.00005$ & $0.056 \pm 0.019$ & $0.389\pm 0.159$ & $5.0$ & $ $ \\
$f_{      60}$    & $0.00915 \pm 0.00005$ & $0.088 \pm 0.032$ & $0.485\pm 0.171$ & $4.7$ & $ $ \\
$f_{      61}$    & $0.01408 \pm 0.00005$ & $0.086 \pm 0.032$ & $0.643\pm 0.174$ & $4.6$ & $ $ \\
$f_{      62}$    & $0.00719 \pm 0.00005$ & $0.085 \pm 0.032$ & $0.115\pm 0.177$ & $4.5$ & $ $ \\
$f_{      63}$    & $7.51321 \pm 0.00005$ & $0.033 \pm 0.013$ & $0.543\pm 0.177$ & $4.5$ & $f_{7}-f_{3}$ \\
$f_{      64}$    & $3.05276 \pm 0.00005$ & $0.039 \pm 0.015$ & $0.643\pm 0.178$ & $4.5$ & $f_1+f_{orb}$ \\
$f_{      65}$    & $2.18949 \pm 0.00005$ & $0.047 \pm 0.018$ & $0.709\pm 0.179$ & $4.5$ & $ $ \\
$f_{      66}$    & $20.00926 \pm 0.00005$ & $0.023 \pm 0.009$ & $0.403\pm 0.180$ & $4.4$ & $ $ \\
$f_{      67}$    & $3.87292 \pm 0.00005$ & $0.032 \pm 0.012$ & $0.336\pm 0.182$ & $4.4$ & $f_1+2f_{orb}$ \\
$f_{      68}$    & $19.67112 \pm 0.00005$ & $0.023 \pm 0.009$ & $0.386\pm 0.183$ & $4.4$ & $ $ \\
$f_{      69}$    & $3.24279 \pm 0.00005$ & $0.036 \pm 0.014$ & $0.543\pm 0.184$ & $4.3$ & $f_3+f_{orb}$ \\
$f_{      70}$    & $0.04100 \pm 0.00005$ & $0.080 \pm 0.032$ & $0.630\pm 0.186$ & $4.3$ & $ $ \\
$f_{      71}$    & $3.50826 \pm 0.00006$ & $0.033 \pm 0.013$ & $0.505\pm 0.189$ & $4.2$ & $ $ \\
$f_{      72}$    & $5.14855 \pm 0.00006$ & $0.025 \pm 0.010$ & $0.339\pm 0.189$ & $4.2$ & $ $ \\
$f_{      73}$    & $10.14854 \pm 0.00006$ & $0.047 \pm 0.019$ & $0.853\pm 0.189$ & $4.2$ & $f_{45}-f_{orb}$ \\
$f_{      74}$    & $14.61031 \pm 0.00006$ & $0.032 \pm 0.013$ & $0.264\pm 0.189$ & $4.2$ & $ $ \\
$f_{      75}$    & $9.84149 \pm 0.00006$ & $0.046 \pm 0.019$ & $0.553\pm 0.193$ & $4.1$ & $ $ \\
$f_{      76}$    & $19.68325 \pm 0.00006$ & $0.022 \pm 0.009$ & $0.630\pm 0.195$ & $4.1$ & $ $ \\
$f_{      77}$    & $9.15351 \pm 0.00006$ & $0.041 \pm 0.017$ & $0.098\pm 0.196$ & $4.1$ & $ $ \\
$f_{      *}$     & $2.53775 \pm 0.00007$ & $0.035 \pm 0.017$ & $0.381\pm 0.224$ & $3.6$ & $l=1,m=1$  \\
\enddata 
\end{deluxetable}


\clearpage



\begin{figure} 
\begin{center} 
 {\includegraphics[angle=90,height=12cm]{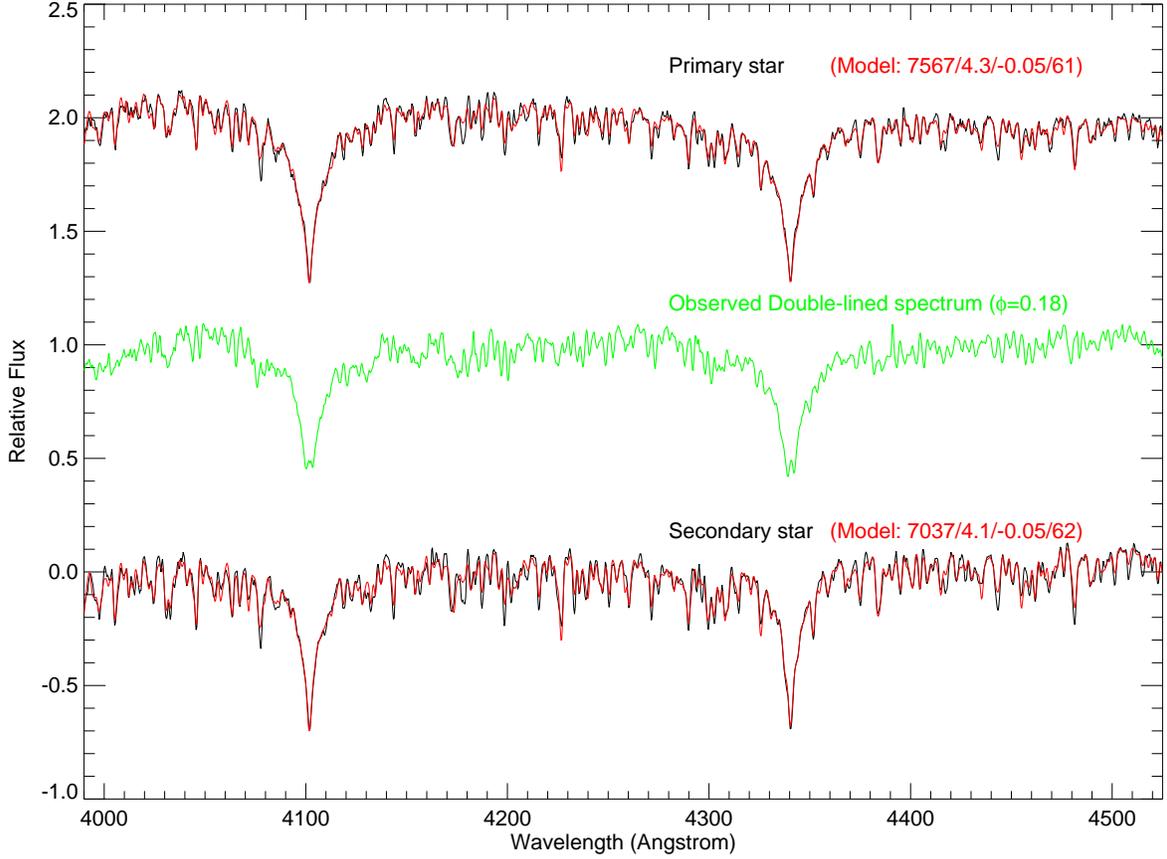}} 
\end{center} 
\caption{The observed composite spectrum at orbital phase $\phi=0.18$ is shown as the green line. The tomographic reconstructed individual spectrum of the primary and secondary star are indicated by the black lines (shifted \textbf{upward and downward} respectively for clarity), with the best-fitting model from UVBLUE over-plotted in red. The atmospheric parameters of the models are shown in the parentheses, in the order of $T_{\rm eff}$(K)/ $\log g(cgs)$/ [Fe/H](dex)/ $v\sin i$(km s$^{-1}$).}
\end{figure}

\begin{figure} 
\begin{center} 
 {\includegraphics[angle=0,height=12cm]{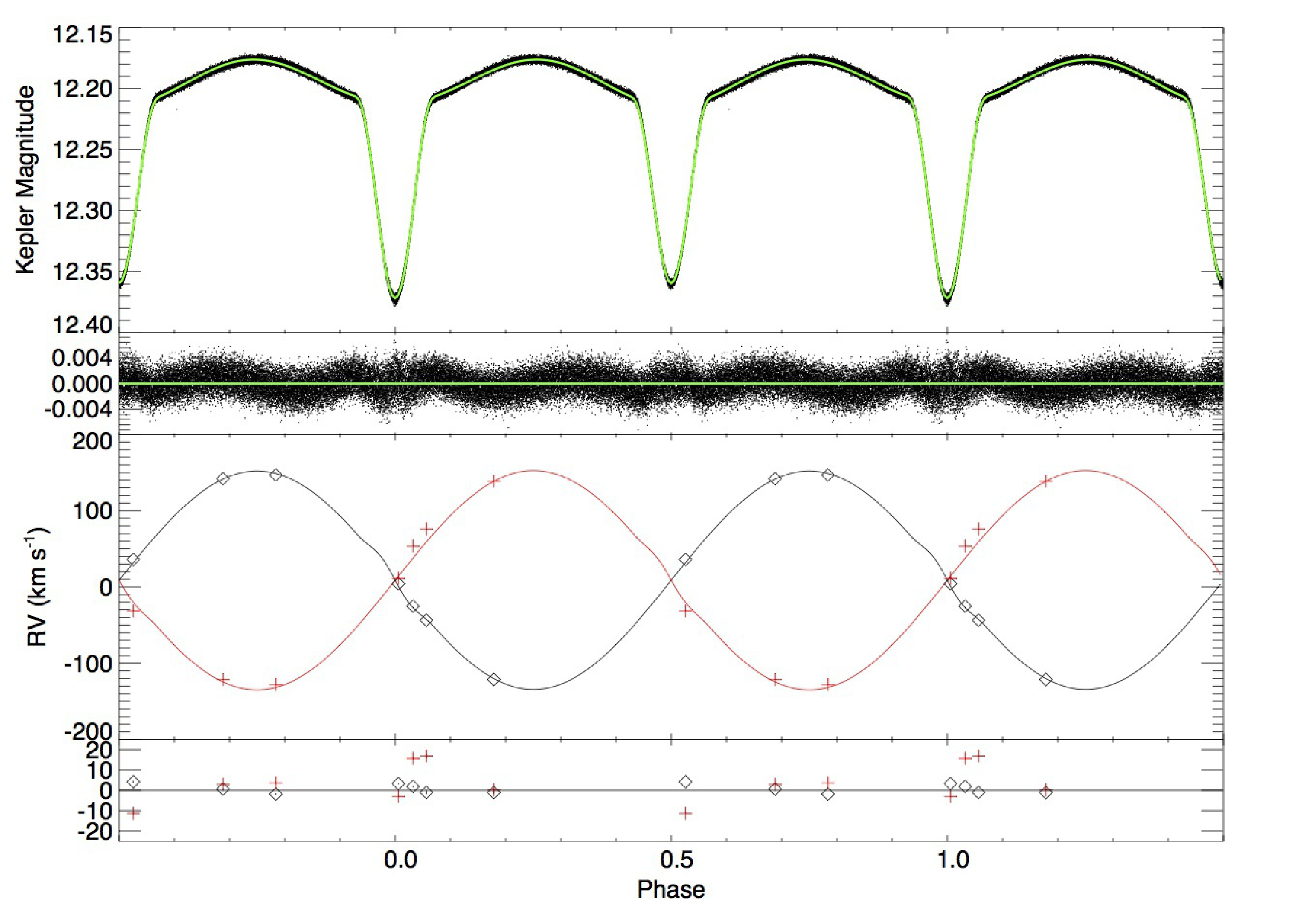}} 
\end{center} 
\caption{The upper and lower panels show the {\it Kepler} light curve (black dots) and radial velocities derived by Matson et al.\ (2017), respectively.  The measured RVs of the primary and the secondary star are indicated by the diamonds and crosses, respectively. The best-fitting binary models from ELC code are over-plotted, with the green line indicating the light curve model and black/red lines for the radial velocity models.}
\end{figure}

\begin{figure} 
\begin{center} 
 {\includegraphics[angle=90,height=13cm]{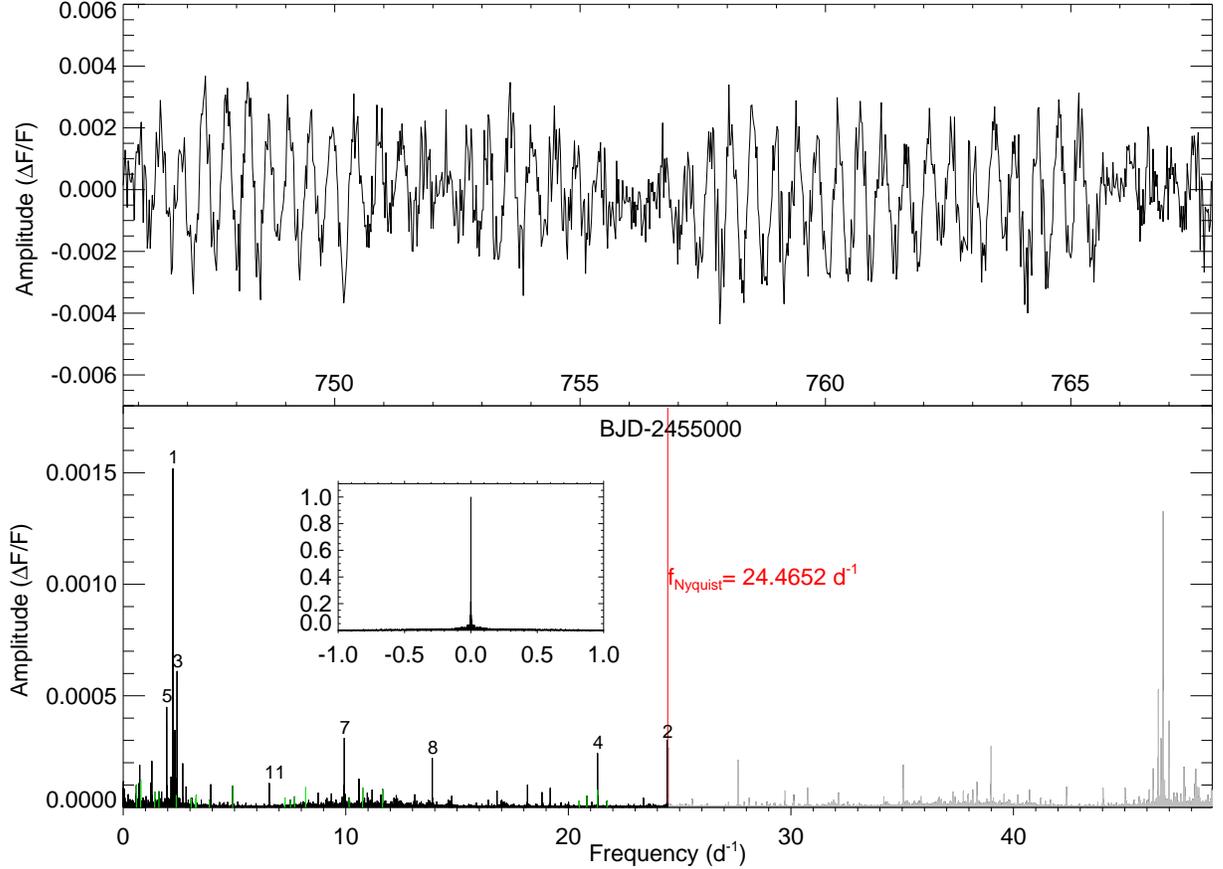}} 
\end{center} 
\caption{Time series of {\it Kepler} photometric data of oscillations with the binary light curve subtracted in the time domain (upper) and frequency domain (lower). Several of the strongest pulsation peaks have been labeled by their frequency numbers (Table 3 and 4). The inset in the lower panel shows the spectral window. The Fourier spectrum in the super-Nyquist region is also shown, with the red vertical line indicating the Nyquist frequency $f_{\rm Nyquist}=24.4652$ d$^{-1}$ of {\it Kepler} long cadence data. All pulsations in the super-Nyquist region (gray shaded) have lower amplitudes compared to their mirror reflections around the $f_{\rm Nquist}$, indicating they are not real pulsation peaks, but are aliases.}
\end{figure}

\begin{figure} 
\begin{center} 
{\includegraphics[height=13cm]{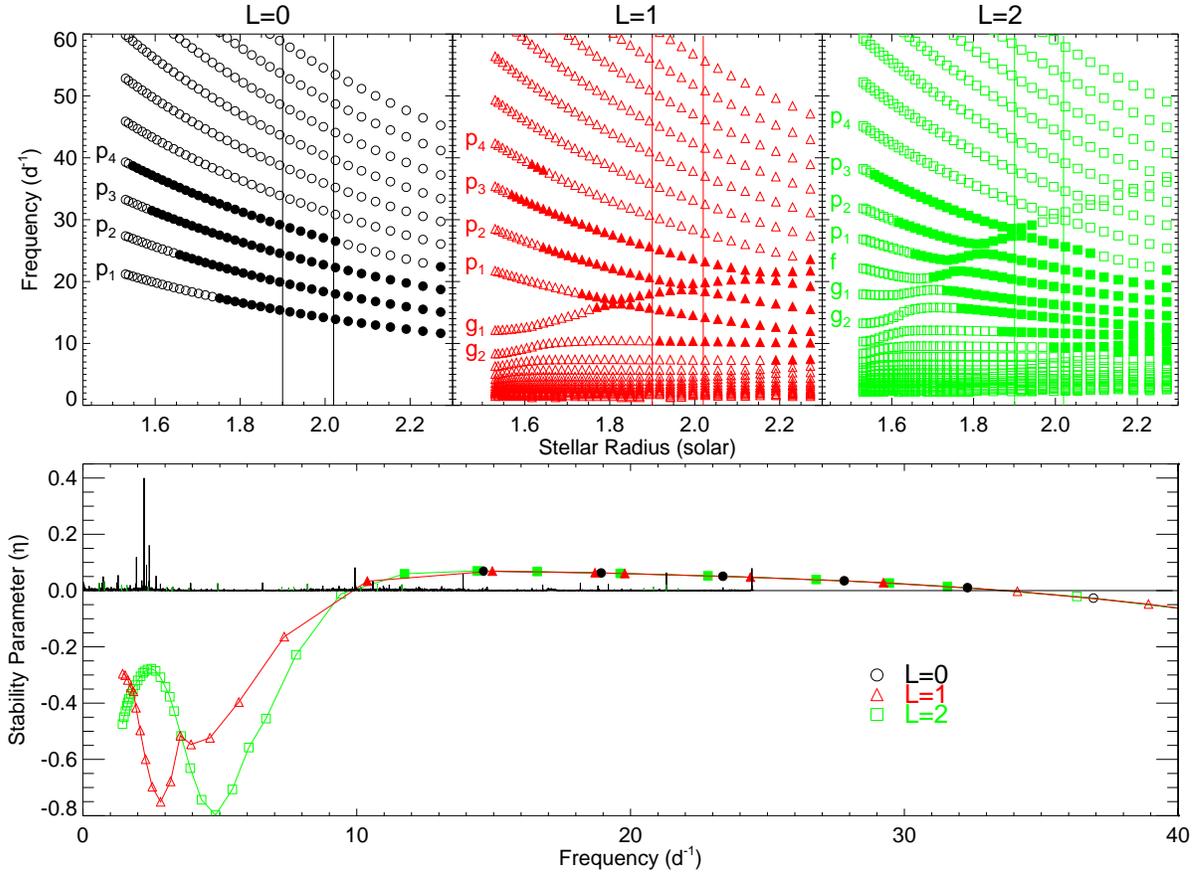}} 
\end{center} 
\caption{\textbf{Upper panel}: The evolution of theoretical pulsation frequencies of $l=0, 1, 2$ modes (black circles, red triangles, and green squares, respectively) of a stellar model with $M=1.70M_{\odot}$. The evolution is shown from ZAMS ($R=1.5R_{\odot}$) to near TAMS ($R=2.3R_{\odot}$). The observed radius of the secondary star ($R=1.96\pm 0.06 R_{\odot}$) is enclosed by the two vertical lines in each sub-panel. The open and filled symbols represent the stable and unstable modes, respectively. \textbf{Lower panel:} The stability parameter $\eta$ of the pulsation modes of the best-matching model with a radius of $R=1.96R_{\odot}$. For clarity, the observed Fourier spectrum is over-plotted and scaled to have a maximum amplitude of 0.4. Combination frequencies are highlighted green.}
\end{figure}

\begin{figure} 
\begin{center} 
 {\includegraphics[angle=0,height=13cm]{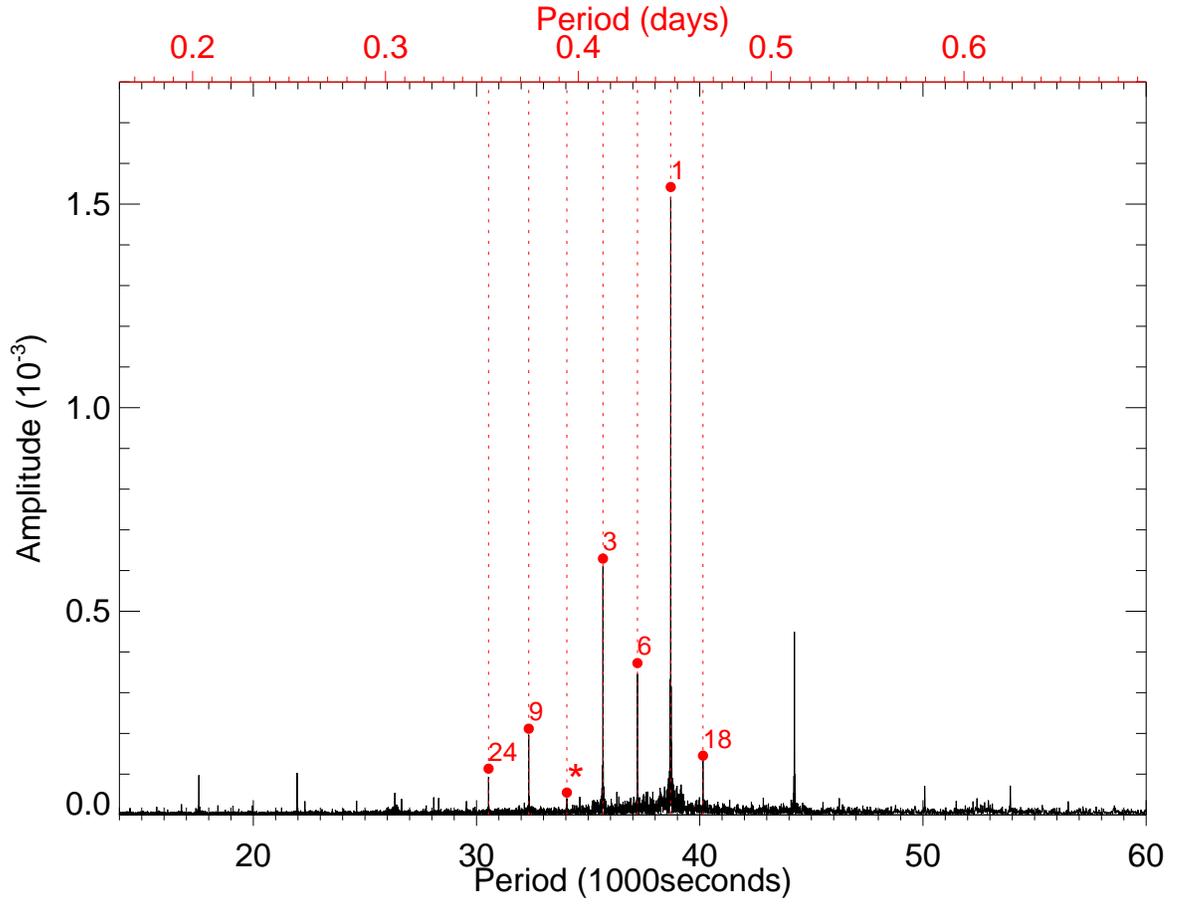}} 
\end{center} 
\caption{The Fourier amplitude spectrum of the g-mode region. The red dotted lines mark the identified series of dipole modes. }
\end{figure}

\begin{figure} 
\begin{center} 
{\includegraphics[height=12cm]{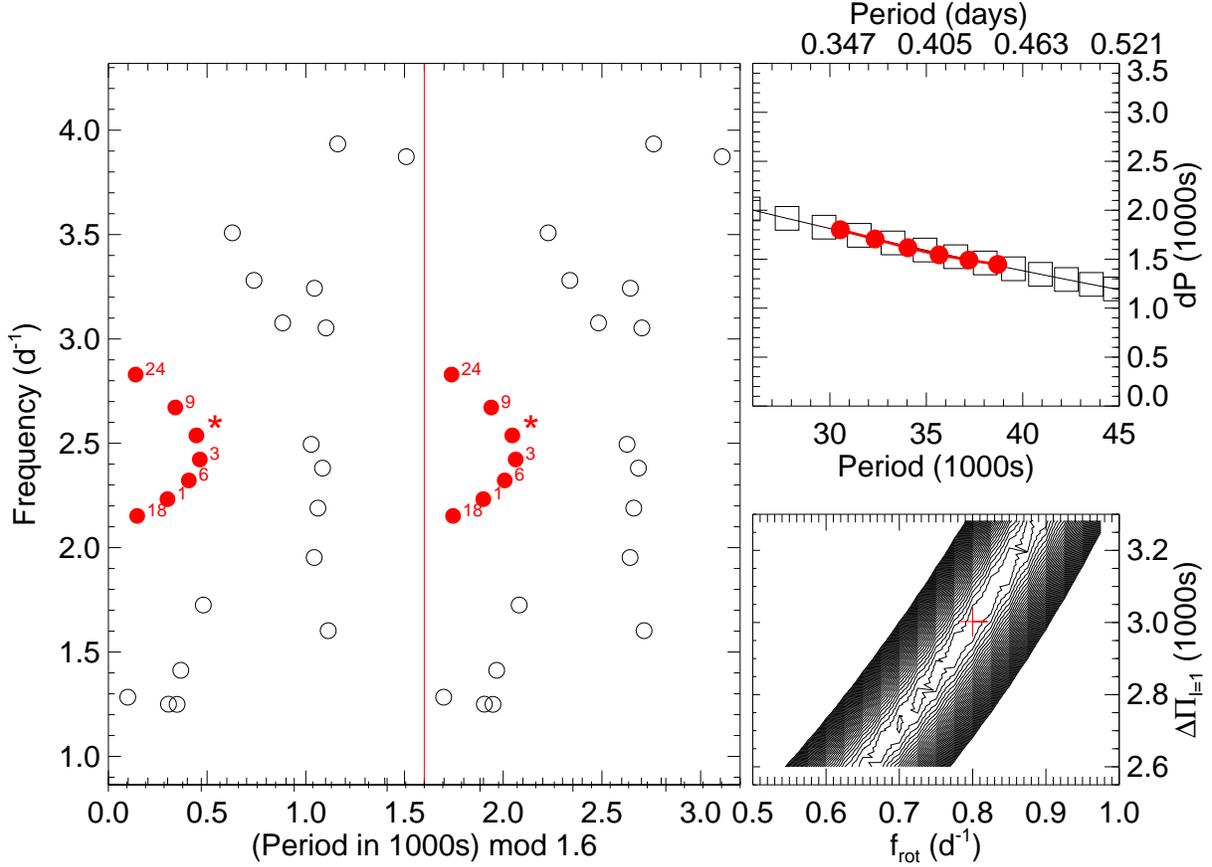}} 
\end{center} 
\caption{\textbf{Left}: The period echelle diagram of observed pulsation modes, plotted twice for clarity. Red symbols represent the identified prograde dipole g-modes around $\sim 2.2$ day$^{-1}$. Their corresponding frequency numbers in Table 3 are labeled. \textbf{Upper right}: The observed period spacing vs. period ($dP - P$) diagram for the identified $l=1,m=1$ g-modes (red symbols). The best-fitting model from the asymptotic period relations in the traditional approximation (eq. 3) is shown as open squares (see text). \textbf{Lower right}: The $\chi^2$ contour from fitting the observed dP $-$ P.  The innermost contour indicates the $1\sigma$ credible region. The best solution ($\Delta\Pi_{l=1}=3000$ sec, $f_{\rm rot}=0.8$ d$^{-1}$) is marked by the red cross. }
\end{figure}

\begin{figure} 
\begin{center} 
 {\includegraphics[angle=0,height=12cm]{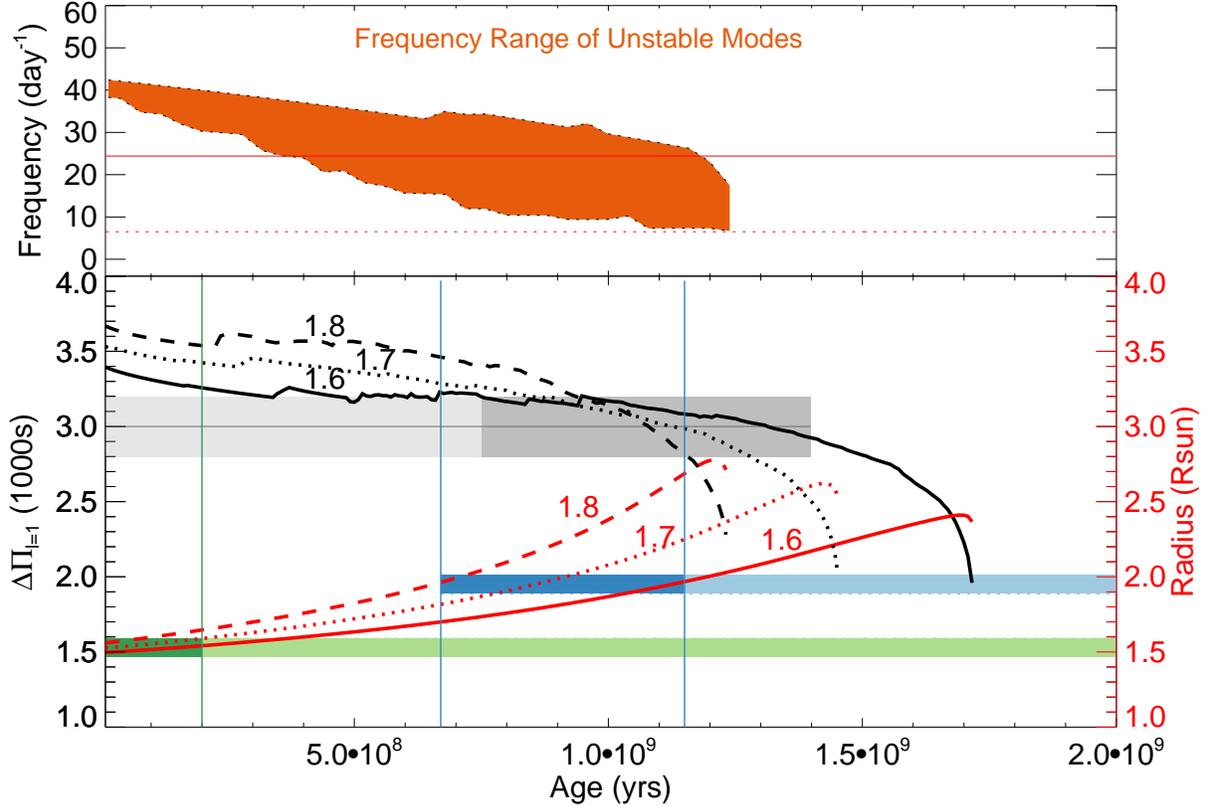}} 
\end{center} 
\caption{\textbf{Upper panel:} The orange shaded region indicates the frequency range of unstable modes ($l=0, 1, 2$) with $\eta >0$ for a $M=1.70 M_{\odot}$ model with solar metalliity. The solid and dotted horizontal lines are the observed maximum frequency and minimum frequency, respectively (when igoring low-frequency g-modes since our code cannot model their excitation).  \textbf{Lower panel:} Evolutionary tracks of the stellar radius (red lines) and the asymptotic dipolar mode period spacing $\Delta\Pi_{l=1}$ (black lines). Tracks with masses of $1.6M_{\odot}, 1.7M_{\odot}$ and $1.8M_{\odot}$ are shown as solid, dotted, and dashed lines, respectively. The green and blue horizontal bars indicate the observed stellar radius of the primary and secondary star, respectively. The observed period spacing of dipolar modes is shown as the gray shaded bar. The deep green, deep blue, and deep gray shaded regions represent the positions on the tracks where models match the observations. The deep gray and deep blue regions overlap in age, suggesting that the observed $l=1,m=1$ modes originate from the secondary star. Coeval models cannot simultaneously explain the observed radius of the primary and the secondary, and this can be seen from the non-overlapping of the deep green and deep blue regions in age.}
\end{figure}

\end{document}